# Adam-Gibbs model in the density scaling regime and its implications for the configurational entropy scaling

E. Masiewicz,[1,2] A. Grzybowski,[1,2,*] K. Grzybowska,[1,2] S. Pawlus,[1,2] J. Pionteck,[3] and M. Paluch[1,2]

[1]*Institute of Physics, University of Silesia, Uniwersytecka 4, 40-007 Katowice, Poland*

[2]*Silesian Center for Education and Interdisciplinary Research, 75 Pułku Piechoty 1A, 41-500 Chorzów, Poland*

[3]*Leibniz Institute of Polymer Research Dresden, Hohe Str. 6, D-01069 Dresden, Germany*

[*] Corresponding author's email: andrzej.grzybowski@us.edu.pl

**ABSTRACT**

To solve a long-standing problem of condensed matter physics with determining a proper description of the thermodynamic evolution of the time scale of molecular dynamics near the glass transition, we extend the well-known Adam-Gibbs model to describe the temperature-volume dependence of structural relaxation times, $\tau_\alpha(T,V)$. We employ the thermodynamic scaling idea reflected in the density scaling power law, $\tau_\alpha = f(T^{-1}V^{-\gamma})$, recently acknowledged as a valid unifying concept in the glass transition physics, to discriminate between physically relevant and irrelevant attempts at formulating the temperature-volume representations of the Adam-Gibbs model. As a consequence, we determine a straightforward relation between the structural relaxation time $\tau_\alpha$ and the configurational entropy $S_c$, giving evidence that also $S_c(T,V) = g(T^{-1}V^{-\gamma})$ with the exponent γ that enables to scale $\tau_\alpha(T,V)$. This important finding has meaningful implications for the linkage between thermodynamics and molecular dynamics near the glass transition, because it implies that $\tau_\alpha$ can be scaled with $S_c$.

# 1. INTRODUCTION

The phenomenon of glass formation is an important and intriguing area of research in condensed matter physics, continuously attracting a lot of attention of both experimentalists and theoreticians. A key problem in this field is to develop the physical model that will be able to describe the evolution of the structural relaxation time, $\tau_\alpha$, or alternatively viscosity, $\eta$, on approaching the glass transition. So far, the most efforts have been concentrated on the analysis and the correct description of the temperature dependence of $\tau_\alpha$ at ambient pressure. In this context, the question is often raised whether the structural relaxation dynamics diverges at some finite temperature. However, the major challenge is to formulate the appropriate equation of state, i.e.: to provide the analytical expression for description of $\tau_\alpha$ in the full (temperature-pressure-volume) thermodynamic space. This challenge is prompted by the fact that the experimental studies of the molecular dynamics of glass-forming systems at elevated pressure are now possible in many laboratories all over the world.

Among variety of models proposed for description of the temperature dependence of $\tau_\alpha$, the entropy-based model formulated by G.Adam and J.H.Gibbs (AG)[1] has become the most popular and meaningful one as reflected in the number of citations (cited more than 4000 times). This seminal work provides a connection between thermodynamic and dynamic quantities. According to the Adam-Gibbs model, the structural relaxation time of supercooled liquid is controlled by the configurational entropy $S_C(T)$ which determines the size of cooperatively rearranging regions (CRR). This has been expressed by the following formula

$$\tau_\alpha = \tau_{AG}\, exp\left(\frac{A}{T\, S_C}\right) \qquad (1)$$

Here $S_C(T)$ is defined as the configurational entropy, $S_C(T) = S^{melt} - S^{solid}$, $A$ is the constant related to the intermolecular potential and is also proportional to the free energy barrier (per molecule in CRR) for rearrangement $\Delta\mu$ while $\tau_{AG}$ is the value of structural relaxation time in the limit of high temperatures. CRRs are defined as the smallest volume elements that can relax to the new configurations independently of their environment.

In the last decade, a lot of interest has been directed toward the analysis of molecular dynamics of supercooled liquids in terms of thermodynamic scaling[2,3,4,5,6,7,8,9,10,11,12,13,14,15,16]. This alternative approach is very appealing due to the possibility of universal description of relaxation phenomena for all supercooled liquids based on the generalized Lennard-Jones potential[17]. According to the thermodynamic scaling, some dynamic quantities can be scaled into a single master curve if they are plotted versus $T^{-1}V^{-\gamma}$ , where *T* is the system temperature, *V* is the system specific volume and $\gamma$ is the scaling exponent[12,13]

$$log_{10}x = f(T^{-1}V^{-\gamma}). \qquad (2)$$

The variable *x* denotes one of the dynamic quantities, such as the structural relaxation time $\tau_\alpha$, viscosity $\eta$, or other dynamic property. The key quantity is here the scaling exponent $\gamma$. At the first stage of development of the thermodynamic scaling approach, it was postulated[18] that the value of scaling exponent should be equal to 4 in accordance with the initial finding for OTP[18,19]. However, it has been subsequently proved by a number of research groups that the value of $\gamma$ can significantly differ from 4 for other glass formers For example, for van der Waals liquids: PDE: $\gamma = 4.5$, BMPC: $\gamma = 7.0$, BMMPC: $\gamma = 8.5$;[11,12,13,14,16,20,21] for polymers: $\gamma = 1.9 \div 5.6$;[22,23,24,25,26] for ionic liquids: $\gamma = 2.25 \div 3.7$;[11,15,27] for substances with hydrogen bonds: sorbitol: $\gamma = 0.13$,[11,12], salol: $\gamma = 5.2$.[11,12,16,20] Major advances in understanding of the molecular basis of thermodynamic scaling and its relation to macroscopic thermodynamic properties of viscous systems have been possible to achieve by performing molecular dynamics (MD) simulations. Assuming that a short range effective intermolecular potential can be approximated by a combination ofdominating repulsive inverse power law and small attractive background, the validity of the thermodynamic scaling was demonstrated on the basis of MD simulations[6,15,28,29,30,31,32,33]

$$U_{eff}(r) = 4\varepsilon(\sigma/r)^{3\gamma_{IPL}} - A_t\,, \qquad (3)$$

where ε, σ are respectively the potential well depth and the finite distance of the zero potential, which are the typical patemeters of the Lennard-Jones potential, and $A_t$ is a small attractive background. Moreover, it was pointed out that the parameter $\gamma_{IPL}$can be identified with the scaling exponent $\gamma$ in the thermodynamic scaling law (Eq. (2)). This straightforward connection between both exponents made the thermodynamic scaling very attractive approach.

In this context, it is natural to ask how the thermodynamic scaling is incorporated into the AG model. Answering this question requires converting the temperature-dependent AG model (AG(T)) to its T-V representation $AG(T,V)$. Furthermore, the generalization of the AG(T) model to T-V variables might be essential for testing its validity in general.

In this paper, we propose an extension of the original AG model [Eq. (1)] to take into account the combined effect of temperature and volume changes on $\tau_\alpha$. Consequently, we aim to verify whether or not the concept of thermodynamic scaling is consistent with Adam-Gibbs model. These considerations lead us to very important implications for the temperature-density scaling rule for the configurational entropy $S_C$ and the well-grounded relation between $\tau_\alpha$ and $S_C$.

## 2. THEORY

The essential point for applying the $AG$ model, presented by Eq.(1), is the necessity to know the form of $S_C$. The temperature-dependent form of $S_C$, originally proposed by Adam and Gibbs, has been expressed by the following equation

$$S_C(T, P \cong 0) = \int_{T_k}^{T} \frac{\Delta C_P}{T'} dT' = S_\infty - \frac{K_P}{T}, \tag{4}$$

where difference of the isobaric heat capacity of the liquid and crystalline (or the glass) phase varies inversely with temperature $\Delta C_P = K_P/T$, which was found for several glass-formers by R. Richert and C. A. Angell[34] by comparing the behaviour of the dielectric relaxation time with the experimental data of the configurational entropy. $K_P$ is a constant parameter, $T_k$ is Kauzmann's temperature and $S_\infty = K_P/T_k$ is the limiting value of $S_C$ at very high temperatures.

The $AG(T)$ expression (Eq.(1) can be also derived by considering both temperature and pressure dependence of $S_C$. The configurational entropy decreases on cooling or with an increase in pressure, thus the consideration of the dependence of $S_C$ also on pressure, not only on temperature, is very essential. This problem was investigated by Casalini *et al.*[35] by adding the term $S^{isoth}$ to Eq. (4), $S_C(T, P) = S^{isobar} \mp S^{isoth}$ , which explicitly involves the isothermal pressure variation of thermal expansion,

$$S_C(T, P) = \int_{T_k}^{T} \frac{\Delta C_P}{T'} dT' - \int_{P_0}^{P} \Delta \left(\frac{\partial V}{\partial T}\right)_{P'} dP'. \tag{5}$$

If the dependence of the configurational entropy on volume and temperature, $S_C(T, V)$ is known, a much more direct way to test the link between $AG(T, V)$ and the thermodynamic scaling theory is to study the molecular dynamics by Eq.(1) converted to its $T - V$ representation.

To determine the $T - V$ version of Eq.(1), we consider the system entropy as a function of temperature and volume, the total differential of which is given as follows

$$dS = \left(\frac{\partial S}{\partial T}\right)_V dT + \left(\frac{\partial S}{\partial V}\right)_T dV. \tag{6}$$

Using the well-known Maxwell's thermodynamic relationship $(\partial S/\partial V)_T = (\partial P/\partial T)_V$ , and $(\partial S/\partial T)_V = C_V/T$, Eq. (6) can be rewritten as

$$dS = \frac{C_V}{T} dT + \left(\frac{dP}{dT}\right)_V dV, \tag{7}$$

which leads to a temperature-volume function for configurational entropy, $S_C(T, V) = S^{isochor} + S^{isoth}$,

$$S_C(T, V) = \int_{T_k}^{T} \frac{\Delta C_V}{T'} dT' + \int_{V_k}^{V} \Delta \left(\frac{\partial P}{\partial T}\right)_{V'} dV', \tag{8}$$

where the first temperature integral is calculated from the difference between the isochoric heat capacities of the melt and the solid (crystal or glass), $\Delta C_V = C_V^{melt} - C_V^{solid}$ , which can be described over a limited range by $K_V/T$, with a constant $K_V$, similarly to its isobaric counterpart.[34] The second volume integral consituting $S_C(T,V)$ is calculated from the difference between the temperature derivatives of pressure of the melt and the solid (crystal or glass),

$$\Delta\left(\frac{\partial P}{\partial T}\right)_V = \left(\frac{\partial P}{\partial T}\right)_V^{melt} - \left(\frac{\partial P}{\partial T}\right)_V^{solid} .$$

Here we assume that the solid part of the difference between the temperature derivatives of pressure is constant and can be regarded as a fitting parameter. The assumed lower limits of the integratals are respectively Kauzmann's temperature, $T_k$ , and the volume at Kauzmann's temperature for the examined material, $V_k$.

The pressure dependence of temperature at a constant volume can be estimated by using an equation of state (EOS)[36]

$$V(T,P) = V_0\left[1 + \frac{\gamma_{EOS}}{B_T(P_0)}(P - P_0)\right]^{-1/\gamma_{EOS}}, \tag{9}$$

where $B_T(P_0)$ is the isothermal bulk modulus at a reference pressure $P_0$ , parameterized by an exponential temperature function as $B_T(P_0) = b_0 \exp[-b_2(T - T_0)]$ , $\gamma_{EOS}$ is a material constant independent of thermodynamic conditions and $V_0 = V(P_0)$ is the volume at the reference pressure parameterized by a quadratic temperature function, $V_0 = A_0 + A_1(T - T_0) + A_2(T - T_0)^2$, where $T_0 = T_g(P_0)$ is the glass transition temperature at $P_0$. In Eq.(9), $A_0$, $A_1$, $A_2$, $b_0$, $b_2$, $\gamma_{EOS}$ are fitting parameters. Defining new quantities $\delta = -b_2 B_T(P_0)/\gamma_{EOS}$ and $\omega = A_1 + 2A_2(T - T_0)$, the melt part of difference of the temperature derivative of pressure is then given by

$$\left(\frac{\partial P}{\partial T}\right)_V^{melt} = \delta\left[\left(\frac{V_0}{V}\right)^{\gamma_{EOS}} - 1\right] + \omega\frac{B_T(P_0)}{V}\left(\frac{V_0}{V}\right)^{\gamma_{EOS}-1}. \tag{10}$$

Consequently, the integral of Eq.(10) takes the following form

$$\begin{aligned}\int_{V_k}^{V}\left(\frac{\partial P}{\partial T}\right)_{V'}^{melt} dV' \\ &= -\delta(V - V_k) + \frac{1}{1-\gamma_{EOS}}\left(V^{1-\gamma_{EOS}} - V_k^{1-\gamma_{EOS}}\right) \\ &\times\left\{\delta\, V_0^{\gamma EOS} + \omega\, B_T(P_0) V_0^{\gamma EOS-1}\right\}.\end{aligned} \tag{11}$$

Inserting Eq.(11) and $S^{isochor} = S_\infty - K_V/T$ into Eq.(8), we find the expression for $S_c(T,V)$,

$$S_C(T,V) = S_\infty - \frac{K_V}{T} + \left\{ -\left[ \left(\frac{\partial P}{\partial T}\right)_V^{solid} + \delta \right](V - V_k) + \frac{1}{1-\gamma_{EOS}}\left(V^{1-\gamma_{EOS}} - V_k^{1-\gamma_{EOS}}\right) \times \left[\delta\, V_0^{\gamma_{EOS}} + \omega\, B_T(P_0) V_0^{\gamma_{EOS}-1}\right] \right\}. \quad (12)$$

Finally, taking into account the classical AG equation (Eq.(1)) and the expression $S_C(T,V)$, we obtain the *AG(T,V)* representation for $\tau_\alpha(T,V)$

$$\tau_\alpha(T,V) = \tau_0\, exp\left( \frac{A}{\left(T - T_{0AG} + T\frac{1}{S_\infty}\left\{ -\left[\left(\frac{\partial P}{\partial T}\right)_V^{solid} + \delta\right](V - V_k) + \frac{1}{1-\gamma_{EOS}}\left(V^{1-\gamma_{EOS}} - V_k^{1-\gamma_{EOS}}\right) \times \left[\delta\, V_0^{\gamma_{EOS}} + \omega\, B_T(P_0) V_0^{\gamma_{EOS}-1}\right]\right\}\right)} \right), \quad (13)$$

where $T_{0AG} = K_V/S_\infty$ and the parameter $A$ is defined as $C_{AG}\,\Delta\mu/S_\infty$, where $C_{AG}$ is a constant.

If the structural relaxation time and the configurational entropy obey the thermodynamic scaling law in the form of the power law density scaling, $\tau = F(TV^\gamma)$ and $S_C = G(TV^\gamma)$, then a consequence of the thermodynamic scaling hypothesis for the elementary activation energy in the material-specific coefficient $A$ of Adam-Gibbs approach [Eq.(1)] is that it is expected to be not a constant but to comply with a power law dependence of volume in the form $A = A(V) \rightarrow A'\, V^{-\gamma}$. The scenario for the volume (or density) dependence of $A$ in the AG equation was postulated by C. Alba-Simionesco *et al.*,[37] but it was not tested. Using the Kob-Andersen binary Lennard-Jones mixture, an explicit simulation tests of $TV^\gamma$-scaling of $S_C$ and $\tau_\alpha$ in terms of the AG model as well as the scaled volume dependent change in $A(V)$ was successfully performed by S. Sengupta *et al.*[38] Following this approximation, we propose the second formula for $\tau_\alpha(T,V)$, which is a modified Eq. (13) by involving the volume contribution to the parameter *A*

$$\tau_\alpha(T,V) = \tau_0 exp\left( \frac{A'\, V^{-\gamma}}{\left(T - T_{0AG} + T\frac{1}{S_\infty}\left\{ -\left[\left(\frac{\partial P}{\partial T}\right)_V^{solid} + \delta\right](V - V_k) + \frac{1}{1-\gamma_{EOS}}\left(V^{1-\gamma_{EOS}} - V_k^{1-\gamma_{EOS}}\right) \times \left[\delta\, V_0^{\gamma_{EOS}} + \omega\, B_T(P_0) V_0^{\gamma_{EOS}-1}\right]\right\}\right)} \right), \quad (14)$$

where the scaling exponent $\gamma$ is computed from standard methods. The AG(T,V) model, in this form, is a good candidate to be a $TV^\gamma$-scaling model.

## 3. RESULTS AND DISCUSSION

In order to verify the equations (13) and (14), we carried out the high pressure dielectric spectroscopy studies of simple van der Waals liquid - *Tributyl-2-acetylcitrate* (TBAC). They were intended to determine the temperature and pressure dependence of structural relaxation times. Dielectric spectra were measured both at isobaric (0.1 and 200 MPa) and isothermal conditions (199.0K, 202.5K, 205.9K, 209.0K, 212.9K, 216.5K, 225.9K and 240.7K) over a wide frequency range from $10^{-2}$ to $10^{6}$ Hz. In Fig. 1a and Fig. 1b, we show a number of representative dielectric loss spectra obtained at various temperatures at ambient pressure and as a function of pressure at constant T=216.5K , respectively. Lowering temperature has a similar effect as increasing pressure, i.e.: in both cases the relaxation peaks moves to lower frequencies. By analyzing all measured spectra, we determined the temperature and pressure dependence of the structural relaxation times, which were calculated from the inverse frequency of the maximum peak position, $\tau_\alpha = (2\pi\, f_{max})^{-1}$. Having determined values of the structural relaxation times in various *T* and *P* conditions, it was possible to construct the 3D plot depicted in Fig. 2.

As a next step toward the experimental verification of the equations (13) and (14), it is necessary to convert the *T-P* data to their *T-V* representation. Therefore, apart from the high pressure dielectric studies, we additionally performed PVT measurements. Fig. 3a displays the experimentally obtained temperature dependences of specific volume $V(T)$ isobars at labeled pressures, in the range of 10 MPa - 200 MPa. The experimental PVT data for TBAC were satisfactorily parameterized by means of the EOS equation of state (Eq.(9) - solid lines) with the following values of its fitting parameters: $A_0 = (0.8685 \pm 0.0004)\ cm^3/g$, $A_1 = (\ 6.95 \pm 0.05) \times 10^{-4}\ cm^3\ K^{-1}/g$, $A_2 = (4.39 \pm 0.16) \times 10^{-7}\ cm^3\ K^{-2}/g$ , $b_0 = (3148.66 \pm 8.32)\ MPa$ , $b_2 = (5.80 \pm 0.02) \times 10^{-3}\ K^{-1}$, $\gamma_{EOS} = 10.09 \pm 0.02$, assuming the reference state at a fixed glass transition temperature $T_0 = 186.06\ K$ at ambient pressure. The value of adjusted $R^2$ is equal to 0.99998. The above set of data enables us to convert $\tau(T,P)$ to $\tau(T,V)$ , and finally to construct 3D or 2D plots of the structural relaxation times versus *T* and *V*, required to perform the test for the validity of the AG(T,V) model. The best 3D numerical fit of $\tau_\alpha(T,V)$ for TBAC data to Eq. (13) was obtained with the well-adjusted coefficient $R^2$ equal to 0.99884 and the values of thefitting parameters: $log_{10}(\tau_0/[s]) = -11.288 \pm 0.089$ , $A = (1354.27 \pm 30.30)\ K$ , $T_{0AG} = (156.50 \pm 0.78)\ K$, $(\partial P/\partial T)^{solid} = (0.15 \pm\ 0.01)\ MPa/K$. The resulting fit to Eq. (13) is depicted in Fig. 4.

It should be stressed that Eq.(13) has, in general, thirteen parameters, but only four of them are free in the fitting procedure. The other parameters were fixed. Their values were earlier established from PVT measurements using the equation of state $(A_0, A_1, A_2, b_0, b_2, \gamma_{EOS}, T_0)$ as well as from the differential scanning calorimetry measurements ($S_\infty$), whereas the value of $V_K$ was calculated from the equation of state at $T_K$ where the temperature $T_K$ was determined from fitting the dielectric isobar at 0.1MPa to the VFT equation, $\tau_\alpha = B/(T - T_{0VFT})$, on the assumption that $T_{0VFT} = T_K$[34]. Here, for TBAC, $T_K = 156.29\ K$ and $V_K(T_K, P_0) = 0.8482\ cm^3/g$. In Fig. 5, we present the temperature dependence of isobaric heat capacity obtained from the differential scanning calorimetry (DSC) measurements with stochastic temperature modulation (TOPEM). Based on this data, we determined the configurational entropy from Eq.(4) (see the inset in Fig. 5), where $\Delta C_P$ is taken as the difference between two linear functions describing respectively temperature behaviour of $C_P^{liquid}$ and $C_P^{glass}$. A fit to $S_C(T)$ according to $S_C(T) = S_\infty(1 - T_K/T)$ (Eq. 7 in Ref.[34]) yields the value of $S_\infty = (0.6314 \pm 0.0007)\, J\, K^{-1} g^{-1}$. We followed the same fitting procedure in case of Eq. (14). In this context, it is worth noting that this equation basically has the same number of free fitting parameters as the previous one, because the additional parameter γ in Eq. (14) was determined from the criterion for the density scaling [39,40]. The value of the scaling exponent γ, required to construct the thermodynamic scaling plot, was determined from the linear regression of $log_{10}T_\tau$ against $log_{10}V_\tau$ at a few constant structural relaxation times (Fig. 3b). As can be seen in Fig. 3b, the best linear fit was achieved for the value $\gamma = 3.17 \pm 0.01$. Using this value of the exponent γ, we constructed the scaling curve by plotting the structural relaxation times versus the product of the temperature *T* and the specific volume *V* raised to the exponent γ. It is obvious from Fig. 3 that all the scaled experimental isobars and isotherms collapse onto a single master curve. This result is in accord with a general observation of the validity of thermodynamic scaling for van der Waals liquids.

The volume dependence of isothermal and isobaric structural relaxation times determined from dielectric measurements and the best fitting curves obtained using Eq. (14) are displayed in Fig. 6, with the well-adjusted coefficient $R^2$ equal to 0.99904 and the following values of the fitting parameters of Eq. (14): $log_{10}(\tau_0/\,[s]) = -10.847 \pm 0.024$ , $A' = (772.68 \pm 2.04)\, Kcm^{3\gamma}/g^{\gamma}$, $T_{0AG} = (156.40 \pm 0.68)\, K$, $(\partial P/\partial T)^{solid} = (0.793 \pm 0.009)\, MPa/K$. As can be seen, a satisfactory agreement between fits and the experimental points has been achieved. The qualities of the fits to Eqs. (13) and (14), , in principle, imply that the *TV*-generalized AG model, represented by both the two equations, provides a satisfactory description of experimental data. A comparison of the values of the adjusted $R^2$ obtained from fitting experimental data to Eq.(13) (Adj. $R^2$=0.99884) and Eq. (14) (Adj. $R^2$= 0.99906) seems to indicate that both the equations lead to the same outcome. Does it indeed mean that both the equations are internally consistent with the thermodynamic scaling concept?

To answer this question we refer to our recent findings reported in Ref. 41 As we pointed out there, one can formulate some general rules for isobaric $\mathrm{m_P^T}$ and isochoric $\mathrm{m_V^T}$ fragilities, i.e.: (i) *compression brings about the decrease in the isobaric fragility* $m_P^T$ and (ii) *the isochoric fragility* $m_V^T$ *is an invariant parameter with pressure.* They are both valid if the density scaling is satisfied. Consequently, on the basis of above rules, we will be able to check the correctness of the derived equations, because if they work the appropriate trend in the fragilities behavior should be reproduced. The isobaric and isochoric fragilities can be defined in the following way

$$m_x = \left.\frac{d log_{10}\tau_\alpha}{d\,(T_g/T)}\right|_{T=T_g, x=const}, \tag{15}$$

where $x$ stands for either $P$ or $V$, depending on the thermodynamic conditions. Analyzing the temperature dependences of the structural relaxation time at constant pressures, we found for TBAC that $m_P$, calculated from Eq.(13), systematically increases with increasing pressure. It is shown in Fig. 7 (solid squares). Similarly, we calculated the isochoric fragility $m_V$ from Eq. (13) and tested it as a function of pressure. The values of $m_V$ were depicted by open square symbols in the same figure. As can be seen, $m_V$ is not a constant, which is in contradiction with the invariant isochoric fragility rule. In fact, $m_V$ appears to be continuously increasing with increasing pressure, giving a value range of $m_V$ varying from 57.82 to 61.38. Thus, the AG(T,V) model, represented by Eq. (13), exhibits patterns of behavior for the pressure dependences of $m_P$ and $m_V$ which are not coherent with those observed commonly for simple van der Waals liquids. On the other hand, solid circles in Fig. 7 represent the dependence $m_P(P)$ obtained from Eq. (14). The value $m_P$ decreases significantly in the experimental pressure range from $m_P$= 86.32 at ambient pressure to $m_P$ = 80.15 at $P$ = 200 MPa, which agrees with the general trend found in case of van der Waals liquids. In addition, we have established that $m_V$ is pressure-independent within error bars and has a constant value equal to 59.54 at investigated pressure range (open circles in Fig. 7).

From the comparison of the isobaric and isochoric fragilities, obtained from Eq.(13) and Eq. (14), we can see that these equations lead to both quantitatively and qualitatively different results. The above analysis unambiguously shows that the appropriate form of AG-model transformed to the T-V thermodynamic space is that given by Eq. (14), which complies with the following compact representation

$$\tau_\alpha(T,V) = \tau_{AG}\, exp\left(\frac{A'}{TV^\gamma\, S_C(T,V)}\right). \tag{16}$$

It should be stressed that the AG(T,V) model expressed by Eq. (14) has turned out to be consistent with the thermodynamic scaling idea, although Eq. (14) has been derived without any scaling assumptions for the configurational entropy $S_C(T,V)$, because the latter has been employed in Eq. (16) by using Eq. (12). This finding provokes a subsequent important question concerning the scaling of the configurational entropy. According to this, $S_C$ calculated from Eq. (12) should be possible to collapse onto a single curve by plotting it as a function of $TV^\gamma$. Thus, our next step is to check whether or not the configurational entropy $S_C$, similarly to the structural relaxation time $\tau_\alpha$, satisfies the $TV^\gamma$-scaling rule. Prior to doing that, we verify the results given by Eq. (12) with the values of its parameters taken from the fitting experimental dependence $\tau_\alpha(T,V)$ to Eq. (14). In the inset in Fig. 5, we compare the dependence $S_C(T)$ determined from the heat capacity measurements (solid squares) at ambient pressure with that obtained from Eq. (12) (open circles), finding a satisfactory agreement between the dependences $S_C(T)$ determined in these different ways. After this additional confirmation of the validity of Eq. (12), we analyze the temperature and volume dependences of the configurational entropy (see Fig. 8a), and then we plot the dependences $log_{10}T_{S_c}(log_{10}V_{S_c})$ at a few constant $S_C$ (Fig. 8b). As can be seen, these dependences have a linear character. From the simple linear regression, we have found that the value of the slope of all the isoentropic lines ($\gamma_{S_C} = 3.19 \pm 0.02$) is almost equal to the value of the scaling exponent for the structural relaxation time ($\gamma = 3.17 \pm 0.01$). It means that we are able to scale the configurational entropy (see Fig. 8) with the value of the scaling exponent, which very well corresponds to that established for the structural relaxation time ($\gamma_{S_C} \cong \gamma$).

## 4. CONCLUSIONS

The latter finding has very important implications for making a final identification of the role of entropy in the thermodynamic evolution of the time scale of molecular dynamics near the glass transition. An important consequence of the found equivalence of the values of the scaling exponents γ and $\gamma_{S_C}$ for the structural relaxation time $\tau_\alpha$ and the configurational entropy $S_C$ should be a subsequent scaling of $\tau_\alpha$ vs $S_C$. As can be seen in Fig. 9, this scaling indeed occurs, because the structural relaxation times of TBAC can be plotted onto a single muster curve as a function of the configurational entropy. This meaningful result clearly shows that the structural relaxation time can be a single variable function of the configurational entropy, $\tau_\alpha = h(S_C)$ , although the more complex formula (Eq. (16)) is required to meet the power law density scaling criterion in terms of the AG model originally based on Eq. (1). An essential impact of the configurational entropy on the thermodynamic evolution of the time scale of molecular dynamics near the glass transition has been anticipated for many years. For instance, Wolynes and coworkers suggested[42,43] a function $\tau_\alpha = h(S_C)$ based on the random first-order transition theory and showed[44] the power law density scaling of $S_C$ using simulation data in a simple model based on the Lennard-Jones potential. To achieve the power law density scaling $S_C = G(TV^{\gamma_{S_C}})$ with $\gamma_{S_C} \cong \gamma$ for real glass formers, Casalini and Roland proposed[45,46,47] an alternative way to calculate $S_C$, which requires determining reference values of $S_C$ along a chosen isochrone $\tau_\alpha = const$. In this paper, for the first time based on experimental data analyses and without making any limiting assumptions for the configurational entropy calculations, we show that both the structural relaxation time and the configurational entropy follow the same pattern of the power law density scaling behavior, which relies on the same value of the scaling exponent γ, i.e., $\tau_\alpha = F(TV^\gamma)$ and $S_C = G(TV^\gamma)$ . In this way, we solve a long-standing problem with determining the proper effect of thermodynamics on molecular dynamics near the glass transition. The found single variable function, $\tau_\alpha = h(S_C)$ , which is a consequence of the density scaling laws $\tau_\alpha = F(TV^\gamma)$ and $S_C = G(TV^\gamma)$, has a decreasing character (see Fig. 9) which implies that a decrease in the configurational entropy straightforwardly causes the dramatic slowdown in the molecular dynamics (reflected in the rapid increase in its time scale) near the glass transition. Thus, the configurational entropy seems to be sufficient to govern the structural relaxation of supercooled liquids without any contributions from additional factors. For comparison, we have very recently established[48] that such an exclusive impact is not made on the structural relaxation by the total system entropy *S* and the excess entropy $S_{ex}$ (defined as the difference between the total system entropy and the entropy of an ideal gas at the same density and temperature), although both *S* and $S_{ex}$ obey the density scaling law. For *S* and $S_{ex}$, the values of the scaling exponents have been found by us to be considerably different from that valid for $\tau_\alpha$ of a given glass former, and consequently the different values of the scaling exponents rationalize the decoupling observed by us between $\tau_\alpha$ and *S* (or $S_{ex}$) and imply that the relation between $\tau_\alpha$ and *S* (or $S_{ex}$) requires supplementing with an additional density factor.[48] In this context, our findings reported herein become especially useful for further investigations, because they suggest a way to formulate an optimal model of the thermodynamic evolution of the time scale of molecular dynamics of supercooled liquids, which is expected to be able to take a form of a single variable function of the configurational entropy $S_c$ or the scaling variable $TV^\gamma$ in the power law density scaling regime.

## ACKNOWLEDGEMENTS

E. M., K.G., A.G., and M.P. are grateful for the financial support from the Polish National Science Center based on Decision No. DEC-2012/04/A/ST3/00337 within the program MAESTRO 2.

# FIGURES

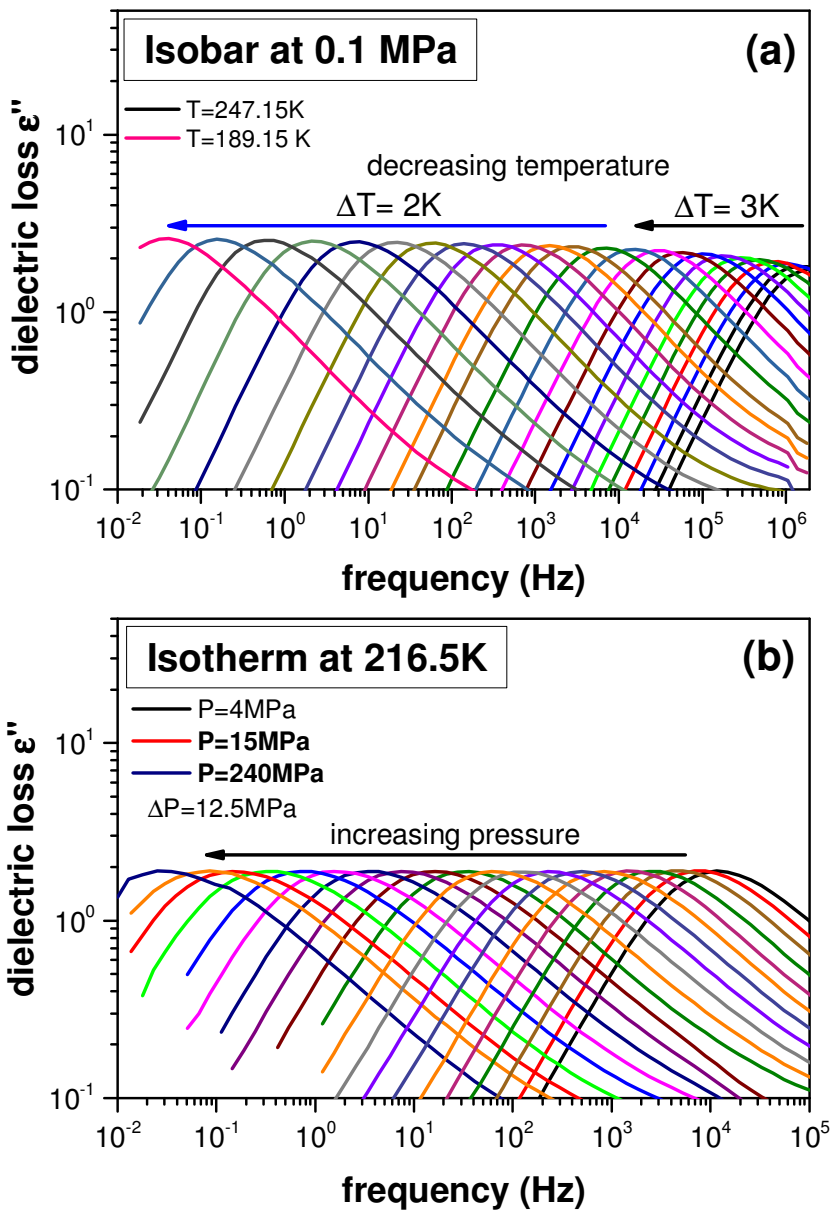

*Fig. 1 Imaginary part of the dielectric loss spectra ε''(ω) for TBAC vs frequency for (a) isobaric measurements at 0.1MPa in the temperature range 189.15K-247.15K in steps of 3K and 2K; (b) for isothermal measurements at 216.5K under increasing pressure.*

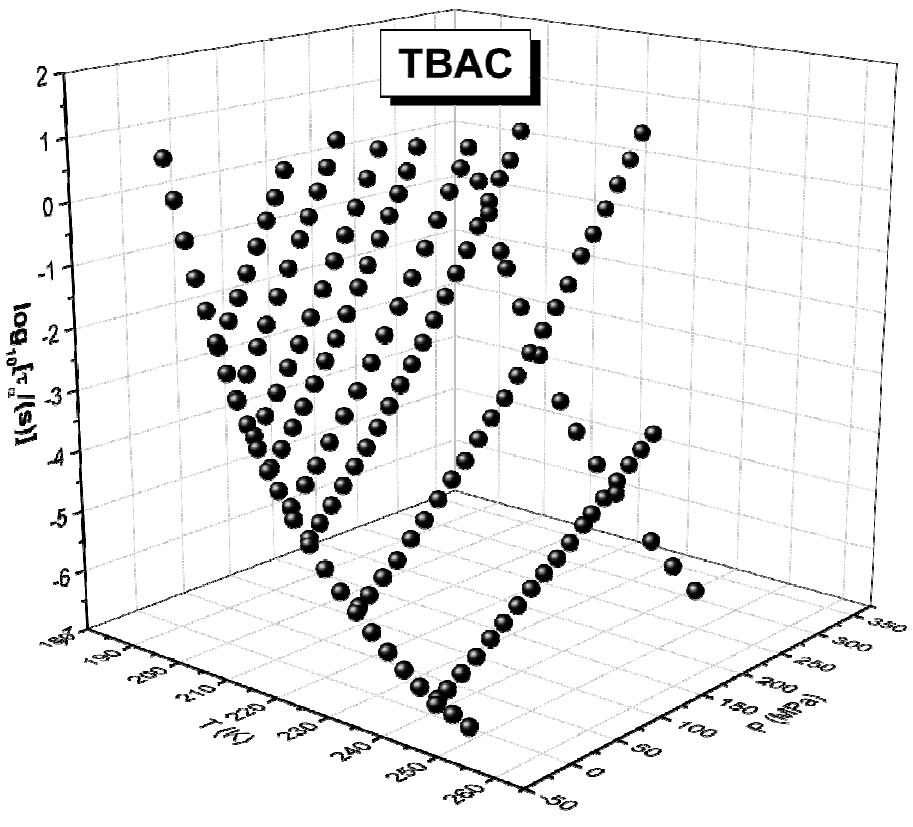

*Fig. 2 Three-dimensional plot of decimal logarithms of isobaric and isothermal structural relaxation times of TBAC as a function of temperature T and pressure P.*

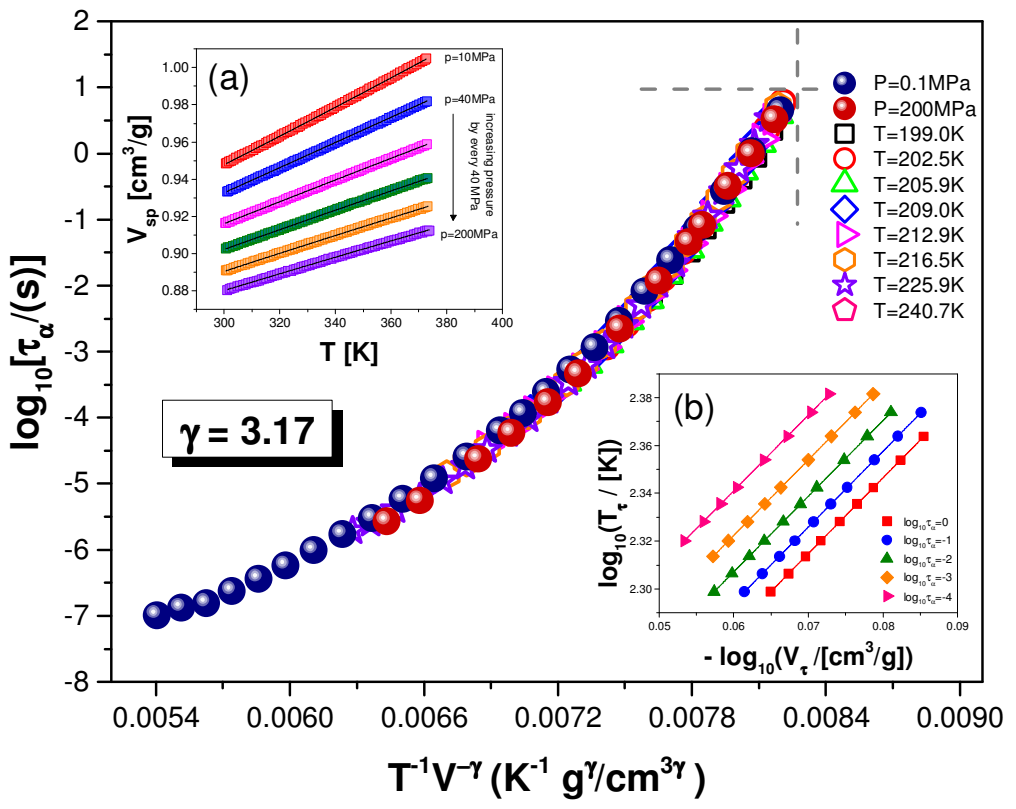

*Fig. 3 Temperature and pressure dependences of structural relaxation times vs. scaling quantity $T^{-1}\,V^{-\gamma}$ with $\gamma = 3.17$. The inset (a) presents isobaric PVT data, $V(T)$. Solid lines are fits to equation of state (EOS) [Eq.(9)]; (b) presents double logarithmic plot of $T_\tau$ versus $V_\tau$ for several relaxation times, as indicated in the inset.*

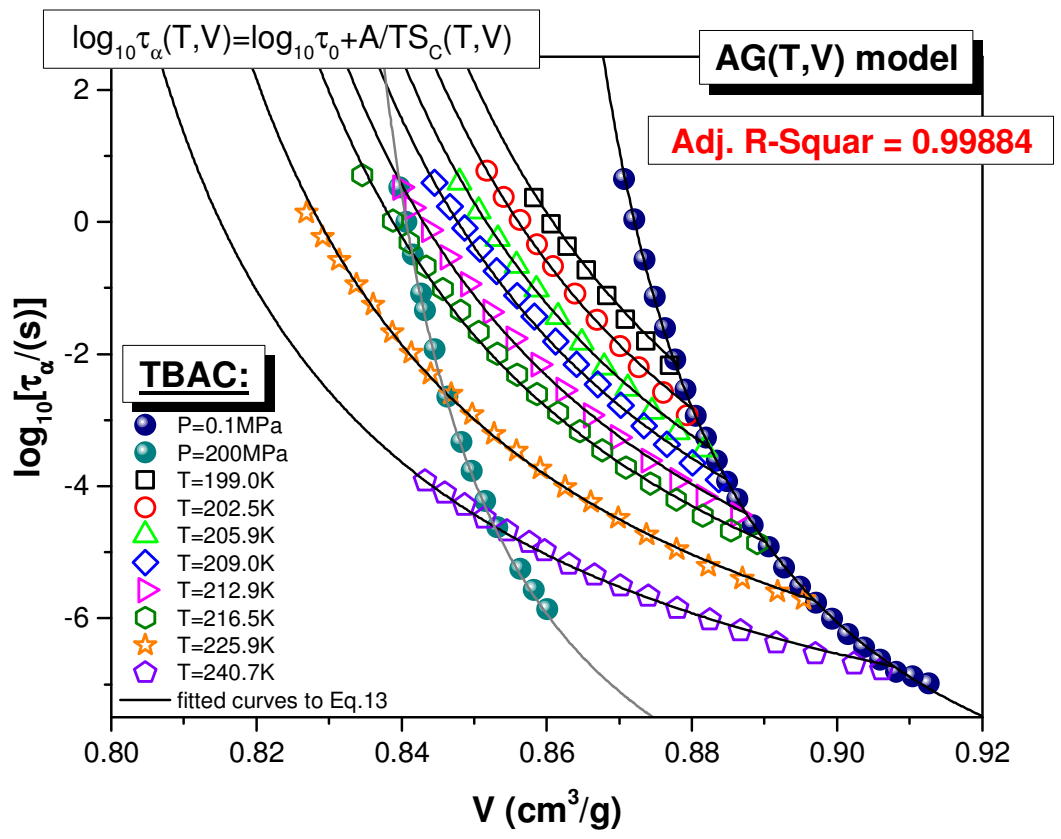

*Fig. 4 The test of AG* $(T,V)$***.*** *The solid lines represent the best fits of* $\tau_\alpha(T,V)$ *to Eq.*(13), projected on the $\tau_\alpha - V$ plane.

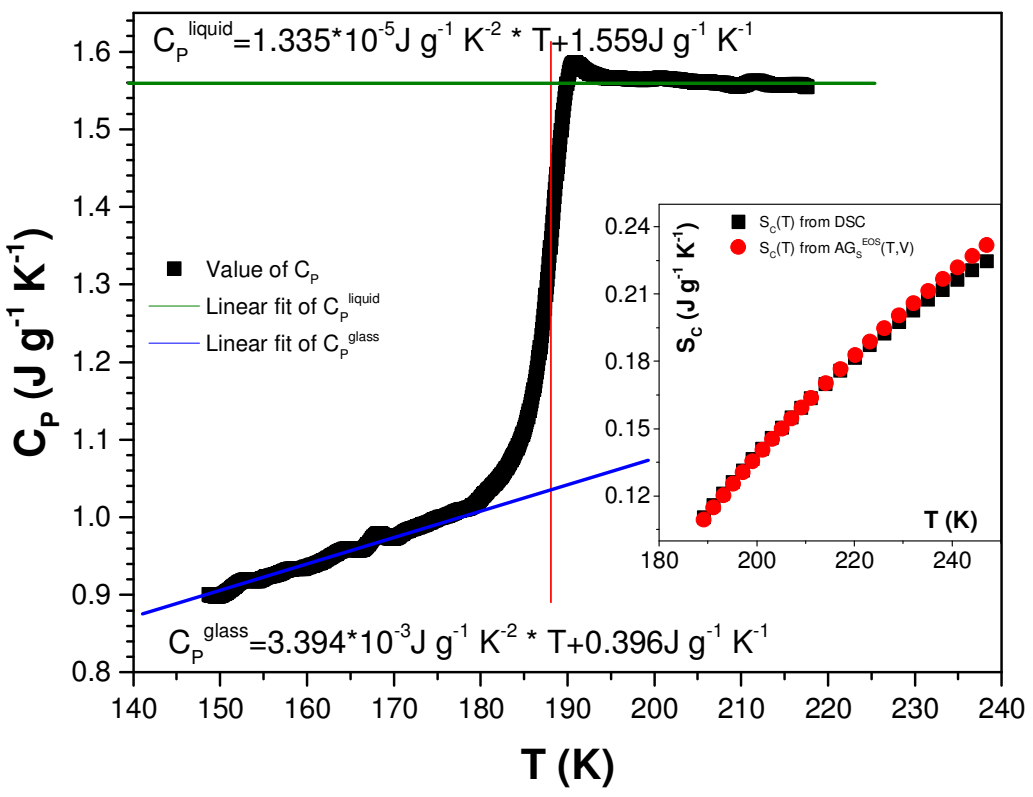

*Fig. 5 Temperature dependences of the heat capacity for TBAC established from TOPEM measurements. The inset shows comparison of the dependence* $S_C(T)$ *determined from the heat capacity measurements at ambient pressure with that calculated from (Eq. 12)* with the values of its paremters found from fitting $\tau_\alpha(T,V)$ to Eq. (14).

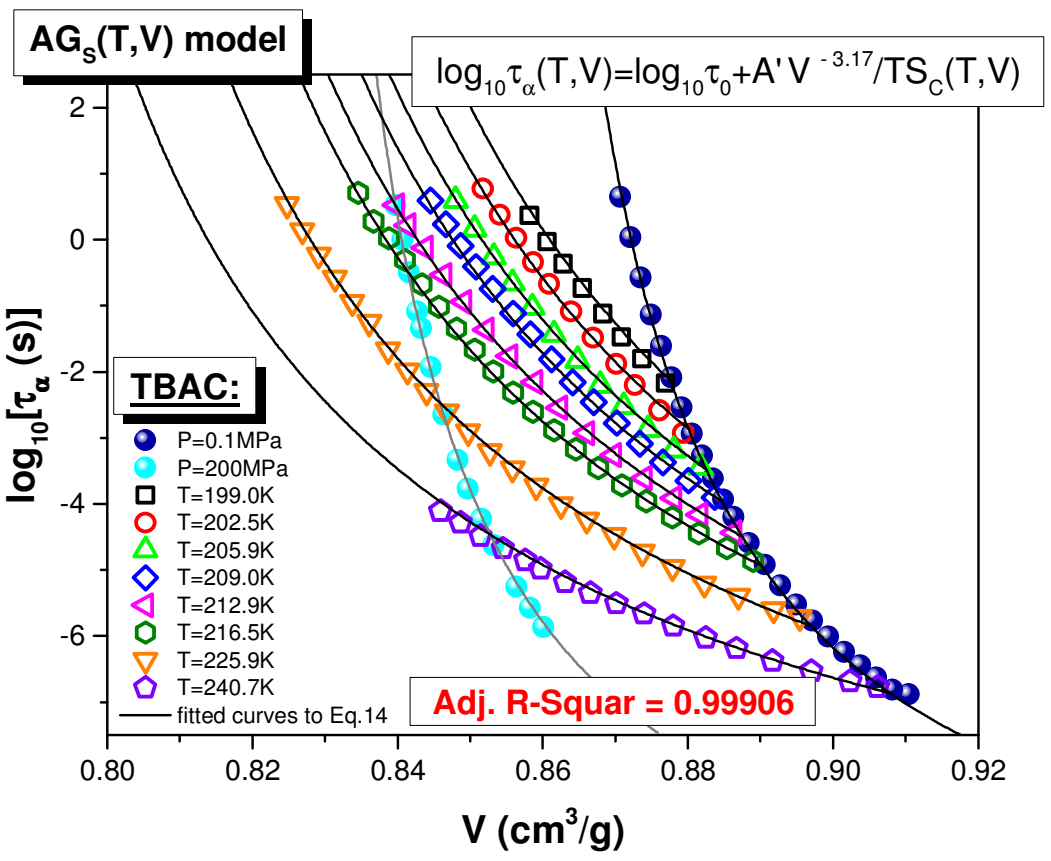

*Fig. 6 Plot of the isobaric and isothermal relaxation data of TBAC vs. specific volume. The solid lines represent the best fits of $\tau_\alpha(T,V)$ to Eq.* (14), projected on the $\tau_\alpha - V$ plane.

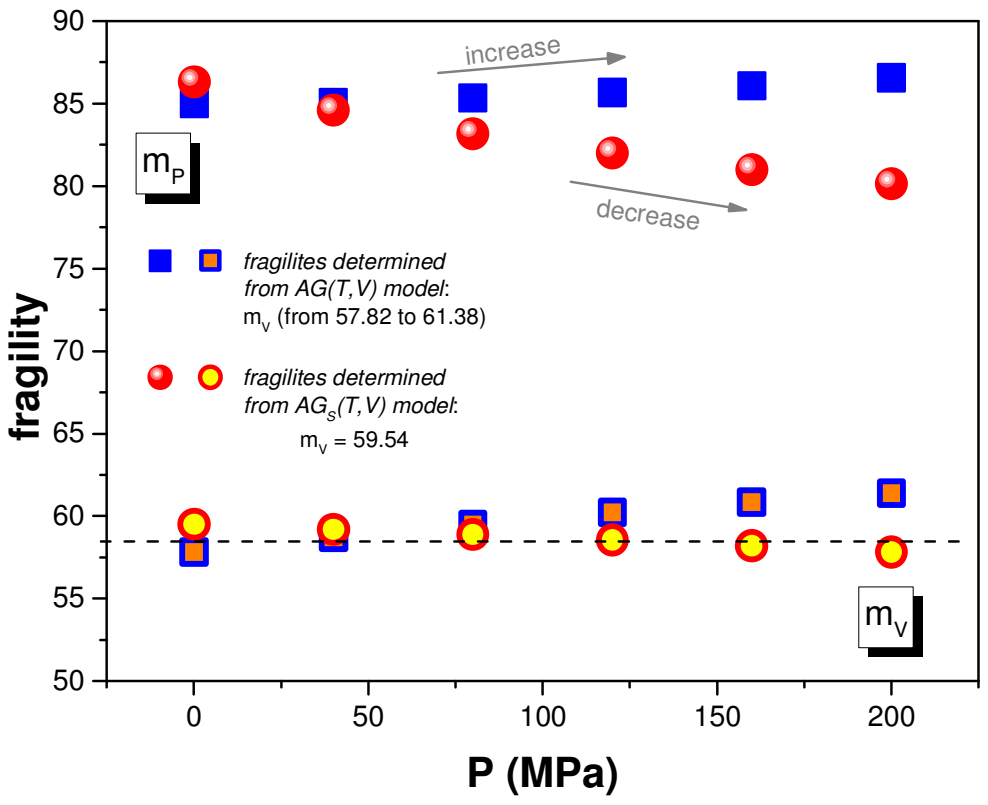

*Fig. 7 Pressure dependences of isobaric fragilities calculated by means of the $AG\ (T,V)$ and $AG_S\ (T,V)$* models (given by Eqs. (13) and (14), respectively) *in the pressure range (0.1 – 200) MPa at $\tau_\alpha = 100s$ and the pressure dependence of isochoric fragilities determined from both the models.*

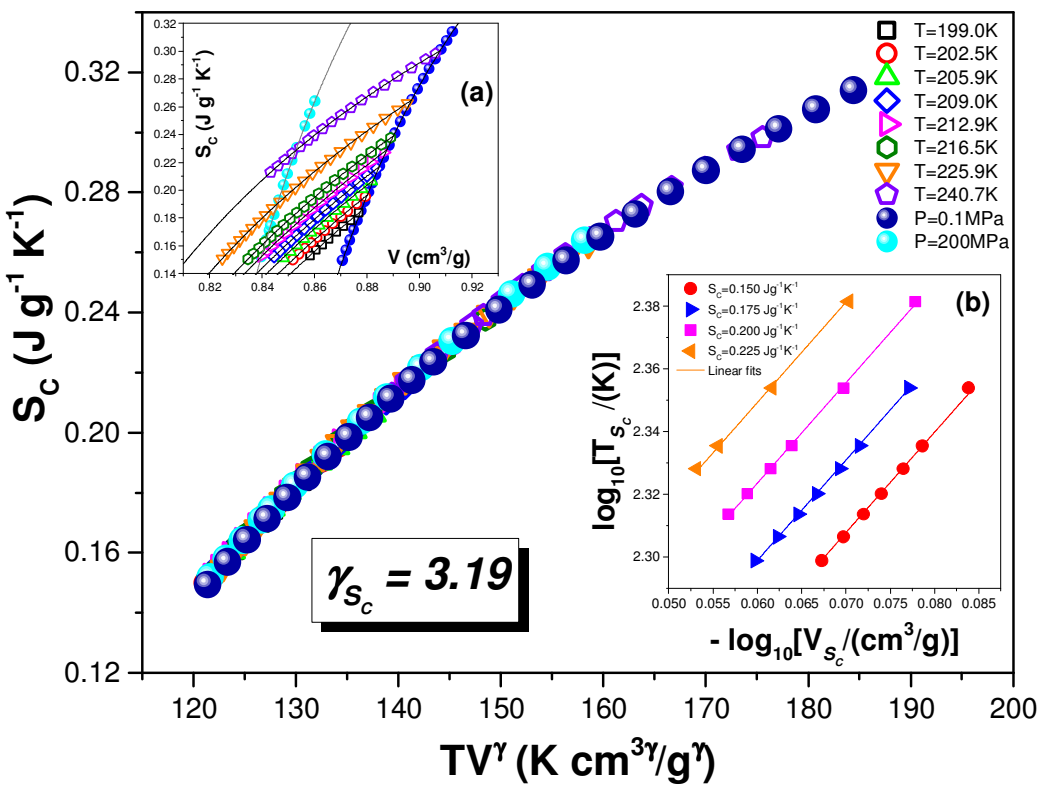

*Fig. 8 Density scaling of the configurational entropy for TBAC vs the scaling quantity* $TV^{\gamma_{Sc}}$ *with* $\gamma_{S_C} = 3.19$*. The inset presents (a) temperature and volume dependences of* $S_C$*; (b) plot of* $log_{10}T_{S_C}(log_{10}V_{S_C})$ *at a few constant* $S_C$*.*

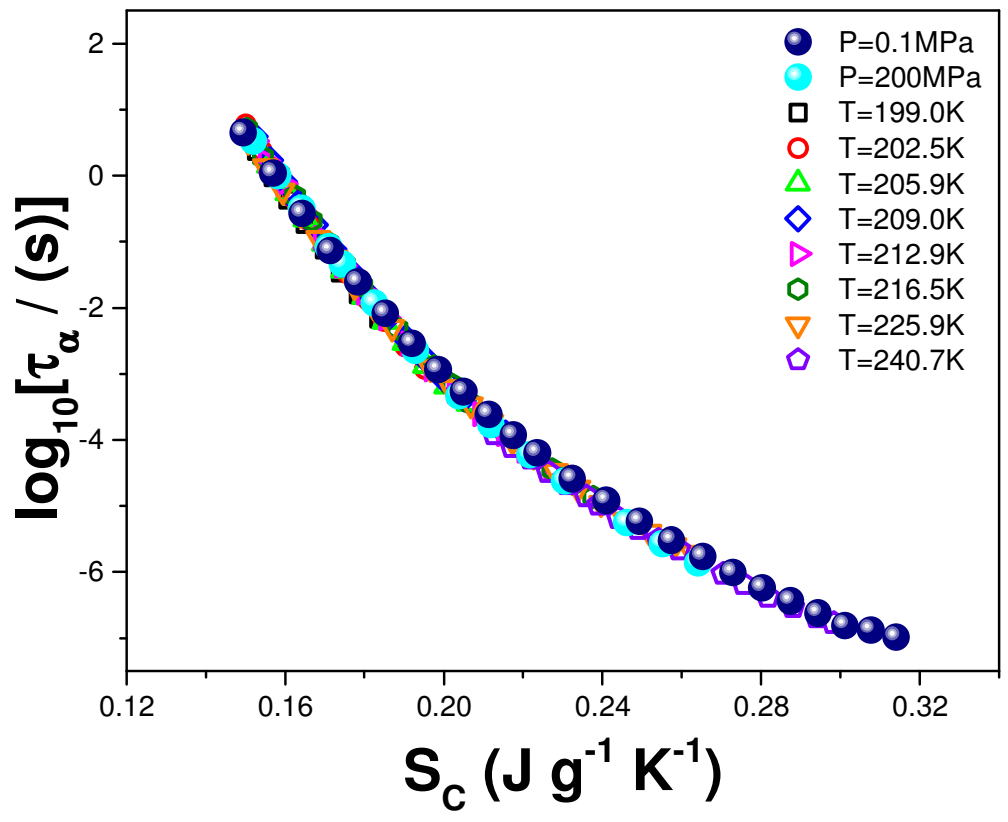

*Fig. 9 Plot of the decimal logarithm of structural relaxation times of TBAC vs the configurational entropy calculated from Eq. (12) with the values of its parameters found by fitting* $\tau_\alpha(T, V)$ *to Eq. (14).*